\documentclass[12pt]{article}
\usepackage{layout,amssymb,amsmath,epsfig}
\begin{document}

\begin{center}
{\Large \textbf{Anisotropic Compact Stars in $f(G)$ Gravity\\[0pt]
}} \medskip

G. Abbas$^{(a)}$\footnote{%
e-mail: ghulamabbas@ciitsahiwal.edu.pk}, D. Momeni$^{(b)}$\footnote{%
e-mail: d.momeni@yahoo.com}, M. Aamir Ali $^{(c)}$\footnote{%
e-mail:mahr.muhammad.aamir@gmail.com}, R. Myrzakulov$^{(b)}$ \footnote{%
e-mail:rmyrzakulov@gmail.com}, S. Qaisar$^{(a)}$ \footnote{%
e-mail:drqaisar@ciitsahiwal.edu.pk}\\[0pt]
$^{a}$ \textit{Department of Mathematics, COMSATS Institute of \\[0pt]
Information Technology Sahiwal-57000, Pakistan}\\[0pt]
$^{b}$\textit{Eurasian International Center for Theoretical Physics and
Department of General Theoretical Physics, Eurasian National University,
Astana 010008, Kazakhstan}\\[0pt]
$^{c}$ \textit{Department of Mathematics, G.C. University, Faisalabad,
Sahiwal Campus,Paksitan}\\[0pt]
\end{center}

\date{}

\begin{abstract}
This paper is devoted to study the possibility of forming
anisotropic compact stars in modified Gauss-Bonnet, namely called as
$f(G)$ theory of gravity which is one of the strong candidates,
responsible for the accelerated expansion of the universe. For this
purpose, we have used analytical solution of Krori and Barua metric
to the Einstein field equations with anisotropic form of matter and
power law model of $f(G)$ gravity. To determine the unknown
constants in Krori and Barua metric, we have used the sample of
compact stars, 4$U$1820-30, Her X-1, SAX J 1808-3658. The physical
behavior of these stars have been analyzed with the observational
data. In this setting, we have checked all the regularity conditions
and stability of the compact stars 4$U$1820-30, Her X-1, SAX J
1808-3658. \newline
\end{abstract}

Keywords: Modified theories of gravity; models beyond the standard models;
compact stars

PACS numbers: 97.60.Jd; 12.60.-i; 04.50.Kd


\section{Introduction}

General relativity (GR) describes many interesting phenomena of
gravity as weak force using the classical gauge theory. Although it
is well stablished both for gravity and to give us an ad hoc model
of Universe, there are two regimes of physics which we need to
modify or motivate new model. The first area is cosmology, the
dynamics of the universe in large scale, where observational data
indicates that we live in an accelerating universe (Riess 1998, 2000
) and (Perlmutter 1998). There is no simple and fully satisfying way
to address this acceleration using the formalism of GR. If we keep
GR, we need to address it to an exotic matter. The only possible and
reasonable way is to modify GR by geometrical corrections, is so
called as modified gravity (Nojiri 2010). Another regime is the
ultraviolet, when we need to modify graviton's propagator
appropriately, specially when we need to present a model for quantum
gravity. In this case, we need to handle a new formalism to cover
Lorentz non-invariance behavior. A way is modification of GR in an
adequate form by adding higher order corrections to the
Einstein-Hilbert action. Different types of modifications for GR
have been proposed in literature: the first and the simplest one was
obtained by replacing the Ricci scalar $R$ by an arbitrary function
$f(R)$, this modification is one of the oldest one and originally
proposed in (Buchdahl 1970) and later revisited by Jedamzik et al.
(2010) and Martin et al. (2014) to address the cosmic acceleration
and also to solve some problems on early universe, specially
inflation. It has been demonstrated (Capozziello and De Laurentis
2011, Nojiri and Odintsov 2011) that $f(R)$ gravity works as a
realistic and physically trustable model.

Curvature corrections can be included in the other forms. For
example one can consider a non minimally model of gravity in the
form of $f(R,L_{m})$ where $L_{m}$ denotes the matter Lagrangian and
$R$ is the scalar curvature (Harko 2010). Another modification is
$f(R,T_{\mu }^{\mu })$ where $T_{\mu \nu }$ is energy-momentum
tensor of matter fields. There are several applications of the above
mentioned models in literature. From string theory point of view,
the next higher order correction to the GR can be induced by
Gauss-Bonnet topological invariant is defined by the following:
\begin{equation*}
G=R^{2}-4R_{\mu \nu }R^{\mu \nu }+R_{\mu \nu \lambda \sigma }R^{\mu \nu
\lambda \sigma }.
\end{equation*}%
Inspired from high energy physics, by looking on GR as the low
energy limit of an affective quantum theory, string theory a valid
modification to the GR was poropsed in which we replace $R$ by
$R+f(G)$ (Nojiri et al. 2005). As we know GB term in four dimensions
has no dynamics so a way is to couple it with the matter field or to
add it by a non linear form $f(G)$. Cosmology of early and late time
universe has been studied as well by GB term (Capozziello 2014 and
Nojiri 2006). Specially exact solutions to this GB-modified gravity
(Nojiri 2000, 2001, 2002) evaporation (Sebastiani et l. 2013) and
cosmic strings (Houndjo et al.  2013 and Rodrigues et al. 2014) have
been investigated in literature.

A class of exact solutions of Einstein field equations with
anisotropic source was presented by Mak and Harko (2004). These
solutions show that energy density as well as tangential and radial
pressure are finite and positive inside the anisotropic star. Chaisi
and Maharaj (2005) formulated an algorithm for the solutions of the
field equations with anisotropic matter source. By using Chaplygin
gas equation of state (EOS), Rahaman et al. (2012) extended the
Krori-Barua (1975) solution to the charge anisotropic source.
Recently (Kalam et al. 2012, Kalam et al. 2013) there has been
growing interest to study the models of the compact objects using
the anisotropic source with Krori-Barua metric. Hossein et al.
(2012) formulated the anisotropic star model with varying
cosmological constant. Bhar et al. (2015) discuss the possibility
for the existence of higher dimensional compact star. Maurya et
al.(2015) formulated the charged anisotropic solutions for the
compact stars.

Compact stars are relativistic massive objects which can be
described by GR as well as modified gravity (Camenzind 2007). Due to
the small size and extremely massive structure they have strong
gravitational force. Recently, there has been growing interest to
study the compact stars in modified theories of gravity like $f(R)$
and $f(G)$ (Astashenok et al. 2014a, 2014b). Specifically, in the
last reference (Astashenok et al. 2014a), the authors presented
neutron stars solutions for viable models of $f(G)$ gravity. In this
paper , motivated by the works of (Momeni et al. 2014) and
(Astashenok et al. 2014a), we study compact stars solutions and
their dynamical stability for a viable power law model of $f(G)$. We
will construct exact solutions for stars which are comparable with
observational data. Our plan in this work is as follows: In
Sec.\textbf{2}, we present basic equations of motion for $f(G)$
gravity. The analytic solution for the viable $f(G)$ model is
presented in Sec.\textbf{3}. Sec.\textbf{4} deals with the physical
analysis of the given system. The last section summaries the results
of the paper.


\section{ Equations of motion of compact stars in modified Gauss-Bonnet
gravity $f(G)$}

\label{eqs} 
To study the dynamics of compact stars in modified GB gravity $f(G)$, we
need to specify our gravitational model. The suitable form of GB action was
presented as the following (Nojiri et al. 2010):
\begin{equation}\label{S}
S=\int d^{4}x\sqrt{-g}\left[ \frac{R}{2\kappa ^{2}}+f(G)\right]
+S_{m}\,\,,
\end{equation}
Where , $G$ is the GB term, $R$ Ricci scalar, $\kappa ^{2}=\frac{8\pi G_{N}}{%
c^{4}}$ denotes gravitational coupling constant, $G_{N}$ the usual Newtonian
constant, $c$ velocity of light in vacuum space time, $S_{m}$ is matter
action. We assume that our GB action given by (\ref{S}) is defined for a
viable form of $f(G)$ which is consistent with the observational data in
accelerating universe,from different observational constraints, like solar
system tests, Cassini experiment and so on. Furthermore, we assume that $f(G)
$ is smooth function of argument $G$ and have all higher order derivatives $%
f^{n\geq 2}(G)$. Using the matter action $S_{m}$, it is possible to define
energy-momentum of matter fields by the standard definition $T_{\mu \nu }=-%
\frac{2}{\sqrt{-g}}\frac{\delta S_{m}}{\delta g^{\mu \nu }}$ . If we take
metric $g_{\mu \nu }$ as dynamical variable, the full set of the equations
of motion derived from (\ref{S}) are written in following form (Nojiri et al. 2010):
\begin{eqnarray}\label{eom}
&&R_{\mu \nu }-\frac{1}{2}Rg_{\mu \nu }+8\Big[R_{\mu \rho \nu \sigma
}+R_{\rho \nu }g_{\sigma \mu }-R_{\rho \sigma }g_{\nu \mu }-R_{\mu
\nu }g_{\sigma \rho }+R_{\mu \sigma }g_{\nu \rho }  \\&&\nonumber
+\frac{R}{2}\left( g_{\mu \nu }g_{\sigma \rho }-g_{\mu \sigma
}g_{\nu \rho }\right) \Big]\nabla ^{\rho }\nabla ^{\sigma
}f_{G}+\left( Gf_{G}-f\right) g_{\mu \nu } =\kappa ^{2}T_{\mu \nu
}\\
\label{eom}
\end{eqnarray}%
here $f_{GG...}=\frac{d^{n}f(G)}{dG^{n}}$ . We use the convention for
curvature tensor by adopting the signature of the metric $g_{\mu \nu }$ as $%
(+---)$ , and consequently covariant derivative for a vector field is
defined by $\nabla _{\mu }V_{\nu }=\partial _{\mu }V_{\nu }-\Gamma _{\mu \nu
}^{\lambda }V_{\lambda }$ and the Riemann tensor is given by $R_{\;\mu \nu
\rho }^{\sigma }=\partial _{\nu }\Gamma _{\mu \rho }^{\sigma }-\partial
_{\rho }\Gamma _{\mu \nu }^{\sigma }+\Gamma _{\mu \rho }^{\omega }\Gamma
_{\omega \nu }^{\sigma }-\Gamma _{\mu \nu }^{\omega }\Gamma _{\omega \rho
}^{\sigma }$ . Matter sector satisfies an extra conservation law $\nabla
^{\mu }T_{\mu \nu }=0$.
\par
To have proceed on stars, we choose the metric of the compact star as static
(time independent) and non rotating , spherically symmetric in normal
coordinates $x^{\mu}=(ct,r,\theta,\varphi)$, as the following form :
\begin{eqnarray}
ds^2=c^2 e^{2\phi}dt^2-e^{2\lambda}dr^2-r^2(d\theta^2+\sin^2\theta
d\varphi^2) .  \label{g}
\end{eqnarray}
We assume that matter sector is filled by a fluid with the non zero
components of the anisotropic energy momentum tensor are given by $%
T_{\mu}^{\nu}=diag(\rho c^2,-p_r,-p_{t},-p_{t})$ where $\rho$ is
energy density, $p_r$ radial pressure and $p_t$ tangential component
of the pressure. If we write down the equations of motion given in
(\ref{eom}), we obtain the following set of independent equations
for $(\mu,\nu)=(ct,ct)$ and $(\mu,\nu)=(r,r) $:
\begin{eqnarray}
&& \frac{1}{r^2}(2r\lambda^{\prime 2\lambda}-1)+8e^{-2\lambda} \Big(%
f_{GG}(G^{\prime \prime }-2\lambda^{\prime }G^{\prime })+f_{GGG}(G^{\prime 2}%
\Big) \Big(\frac{1-e^{2\lambda}}{r^2}-2(\phi^{\prime \prime }+\phi^{\prime
2})\Big)  \notag \\
&&+(Gf_{G}-f)e^{2\lambda}=\kappa^2\rho c^2 e^{2\lambda}  \label{tt}
\end{eqnarray}
\begin{eqnarray}
&&\frac{1}{r^2}(2r\phi^{\prime 2\lambda}+1)-(Gf_{G}-f)e^{2\lambda}=\kappa^2
p_r e^{2\lambda} .  \label{rr}
\end{eqnarray}
Trace of (\ref{eom}) gives us the following auxiliary equation:
\begin{eqnarray}
&&R+8G_{\rho\sigma}\nabla^{\rho}\nabla^{\sigma}f_{G}-4(Gf_{G}-f)=-\kappa^2(%
\rho c^2-p_r-2p_t).
\end{eqnarray}
If we expand it using the metric (\ref{g}), we obtain:
\begin{eqnarray}
&& \Big(\phi^{\prime \prime }+\phi^{\prime 2}-\phi^{\prime }\lambda^{\prime
}+\frac{2}{r}(\phi^{\prime }-\lambda^{\prime })+\frac{1-e^{2\lambda}}{r^2}%
\Big) +8 e^{-2\lambda}\Big(\frac{2\phi^{\prime }}{r}+\frac{1-e^{2\lambda}}{%
r^2}\Big) \notag \\\nonumber &&\times\Big(f_{GG}(G^{\prime \prime
}-2\lambda^{\prime }G^{\prime })+f_{GGG}(G^{\prime 2}\Big)
+4(Gf_{G}-f)e^{2\lambda}=\kappa^2e^{2\lambda}(\rho c^2-p_r-2p_{t}) .
\\ \label{trace}
\end{eqnarray}
If we specify the form of the metric functions, then we are able to
calculate pressures and density using Eqs.(\ref{tt})-(\ref{trace}).

\section{Solutions for a viable $f(G)$ model}
In this section, we shall try to solve (\ref{tt},\ref{rr},\ref{trace}) for a
given viable model of $f(G)$. For this reason, we consider the power law
model $f(G)={\alpha }G^{n}$ proposed in (Cognola et al.  2006), where $\alpha $ is
arbitrary constant and $n>0$. If $n\leq \frac{1}{2}$, the $f(G)$ term
dominates on the Einstein term in the regime of small curvature. We
parameterize the metric as the following (Krori and Barua. 1975):%
\begin{equation*}
2\lambda =Ar^{2},\ \ 2\phi =Br^{2}+C,
\end{equation*}%
where $A$, $B$ and $C$ are arbitrary constants to be evaluated by
using some physical matching conditions(see Table\textbf{\ 1}).
Indeed this solution was treated as a singularity free exact
solution for astatic charged fluid sphere. So our main motivation to
introduce this set of functions for static metric goes back to the
singularity free structure of our extended model for compact star.
It is easy to show that neither densities nor curvature terms become
singular in the vicinity of $r=0$ with this appropriate set of
metric functions.
\par
 Energy density and pressure profiles are given
by the following expressions:

\begin{eqnarray}
p_{r}&&=\frac{e^{-Ar^{2}}}{\kappa^{2}}\left[ -\frac{1}{r^{2}}
(4r^{2B}-e^{Ar^{2}}+1)+\alpha G^{n}(n-1)\right],
\end{eqnarray}
\begin{eqnarray}
\rho &&=-\frac{e^{-Ar^{2}}}{\kappa^{2}r^{2}}(2r^{2}A+e^{Ar^{2}}-1)+2\frac{%
4e^{-2Ar^{2}}}{\kappa^{2}}\Big( \alpha n(n-1)G^{n-3}(G^{^{\prime \prime
}}-2ArG^{^{\prime }}) \\
&&+\alpha n(n-1)(n-2)G^{n-3}(G^{^{\prime }})^{2}\Big)\left( \frac{%
1-e^{Ar^{2}}}{r^{2}}-2B(1-r^{2})\right)+2\frac{\alpha }{2k^{2}}G^{n}(n-1),
\end{eqnarray}
\begin{eqnarray}
p_{t}&&=-\frac{e^{-Ar^{2}}}{2\kappa^{2}r^{2}}(2r^{2}A+e^{Ar^{2}}-1)+\frac{%
4e^{-2Ar^{2}}}{\kappa^{2}}\Big( \alpha n(n-1)G^{n-3}(G^{^{\prime \prime
}}+ArG^{^{\prime }}) \\
&&+\alpha n(n-1)(n-2)G^{n-3}(G^{^{\prime }})^{2}\Big)\left( \frac{%
1-e^{Ar^{2}}}{r^{2}}-2(B-Br^{2})\right)-\frac{3\alpha }{2\kappa^{2}}%
G^{n}(n-1)  \notag \\
&&-\frac{e^{-Ar^{2}}}{\kappa^{2}}(B+B^{2}r^{2}-ABr^{2}+\frac{2}{r}(Br-Ar)+%
\frac{1-e^{Ar^{2}}}{r^{2}})-\frac{4e^{-2Ar^{2}}}{\kappa^{2}}\Big(\frac{2Br}{r%
}  \notag \\
&&+\frac{1-e^{Ar^{2}}}{r^{2}}\Big)(\alpha n(n-1)G^{n-1}(G^{^{\prime \prime
}}-2ArG^{^{\prime }})+\alpha n(n-1)(n-1)G^{n-2}(G^{^{\prime }})^{2})  \notag
\\
&&+\frac{e^{-Ar^{2}}}{2\kappa^{2}}\left[ -\frac{1}{r^{2}}
(4r^{2B}-e^{Ar^{2}}+1)+\alpha G^{n}(n-1)\right].
\end{eqnarray}
Also, the radial and transverse equations of state are defined by the
following equations:
\begin{eqnarray}
w_{r}&&=\frac{p_r}{\rho}\equiv{\frac{e^{-Ar^{2}}}{\kappa^{2}}\left[ -\frac{1%
}{r^{2}} (4r^{2B}-e^{Ar^{2}}+1)+\alpha G^{n}(n-1)\right]} \\
&&\times\Big[-\frac{e^{-Ar^{2}}}{\kappa^{2}r^{2}}(2r^{2}A+e^{Ar^{2}}-1)+%
\frac{8e^{-2Ar^{2}}}{\kappa^{2}}\Big( \alpha n(n-1)G^{n-3}(G^{^{\prime
\prime }}-2ArG^{^{\prime }})  \notag \\
&&+\alpha n(n-1)(n-2)G^{n-3}(G^{^{\prime }})^{2}\Big)\left( \frac{%
1-e^{Ar^{2}}}{r^{2}}-2(B-Br^{2})\right)+\frac{\alpha }{\kappa^{2}}G^{n}(n-1)%
\Big]^{-1},  \notag \\
\end{eqnarray}
\begin{eqnarray}
w_{t}&&=-\frac{e^{-Ar^{2}}}{2\kappa^{2}r^{2}}(2r^{2}A+e^{Ar^{2}}-1)+\frac{%
4e^{-2Ar^{2}}}{\kappa^{2}}\Big( \alpha n(n-1)G^{n-3}(G^{^{\prime \prime
}}+ArG^{^{\prime }}) \\
&&+\alpha n(n-1)(n-2)G^{n-3}(G^{^{\prime }})^{2}\Big)\left( \frac{%
1-e^{Ar^{2}}}{r^{2}}-2(B-Br^{2})\right)-\frac{3\alpha }{2\kappa^{2}}%
G^{n}(n-1)  \notag \\
&&-\frac{e^{-Ar^{2}}}{k^{2}}(B+B^{2}r^{2}-ABr^{2}+\frac{2}{r}(Br-Ar)+\frac{%
1-e^{Ar^{2}}}{r^{2}})-\frac{4e^{-2Ar^{2}}}{\kappa^{2}}\Big(\frac{2Br}{r}
\notag \\
&&+\frac{1-e^{Ar^{2}}}{r^{2}}\Big)\Big(\alpha n(n-1)G^{n-1}(G^{^{\prime
\prime }}-2ArG^{^{\prime }})+\alpha n(n-1)(n-1)G^{n-2}(G^{^{\prime }})^{2}%
\Big)  \notag \\
&&+\frac{e^{-Ar^{2}}}{2\kappa^{2}}\left[ -\frac{1}{r^{2}}
(4r^{2B}-e^{Ar^{2}}+1)+\alpha G^{n}(n-1)\right]  \notag \\
&&\times\Big[-2\frac{e^{-Ar^{2}}}{2\kappa^{2}r^{2}}(2r^{2}A+e^{Ar^{2}}-1)+2%
\frac{4e^{-2Ar^{2}}}{\kappa^{2}}\Big( \alpha n(n-1)G^{n-3}(G^{^{\prime
\prime }}-2ArG^{^{\prime }})  \notag \\
&&+\alpha n(n-1)(n-2)G^{n-3}(G^{^{\prime }})^{2}\Big)\left( \frac{%
1-e^{Ar^{2}}}{r^{2}}-2(B-Br^{2})\right)+2\frac{\alpha }{2\kappa^{2}}%
G^{n}(n-1)\Big]^{-1},  \notag \\
\end{eqnarray}
 The behavior of density, radial and transverse pressures and
equation of state (EoS) parameters are given in figures
\textbf{1-5}. In this case EoS parameters have values
$0<{\omega}_t<1$ and $0<{\omega}_r<1$, which shows the fact that
star consists of ordinary matter and effect of $f(G)$ model in the
present setup. We would like to mention that for the plotting of
graphs the values of $A$ and $B$ have been taken from the
table\textbf{\ 1}.

\begin{figure}
\center\epsfig{file=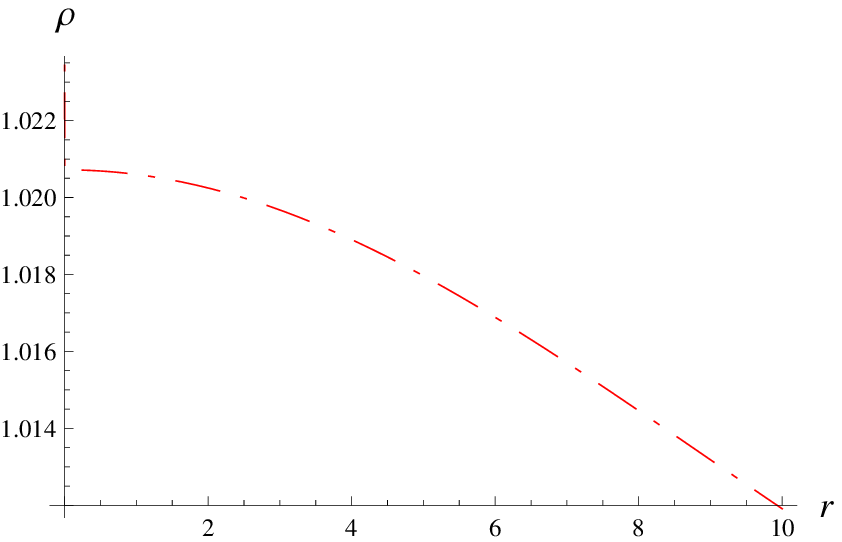, width=0.6\linewidth}
\epsfig{file=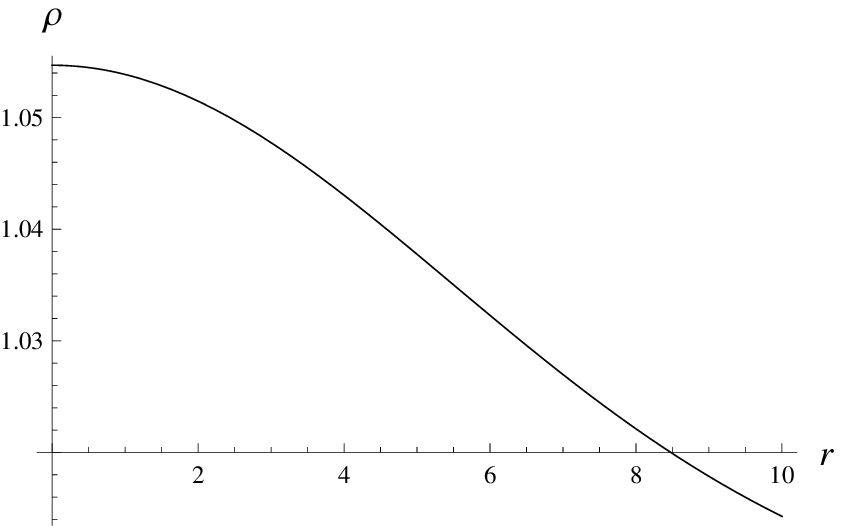, width=0.6\linewidth} \epsfig{file=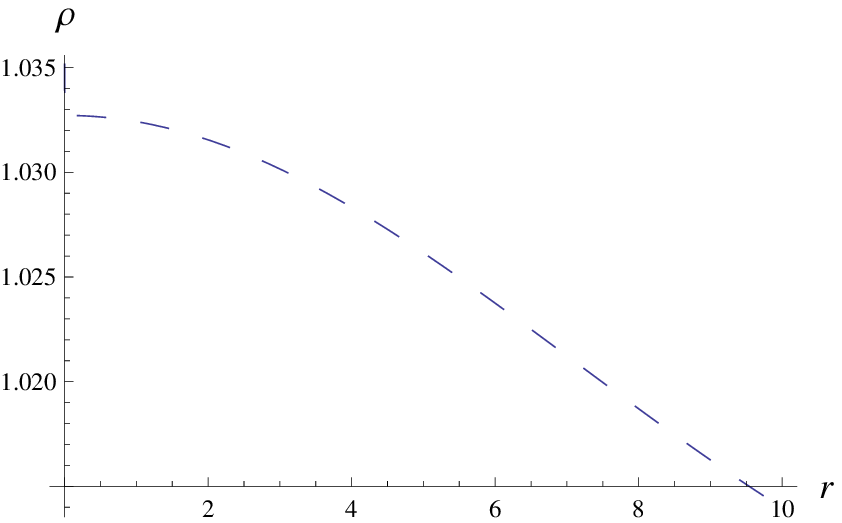,
width=0.6\linewidth}\caption{First, second and third graphs
represent the density variation of Strange star candidate Her X-1,
SAX J 1808.4-3658(SS1) and 4U 1820 - 30, respectively.}
\end{figure}

\begin{figure}
\center\epsfig{file=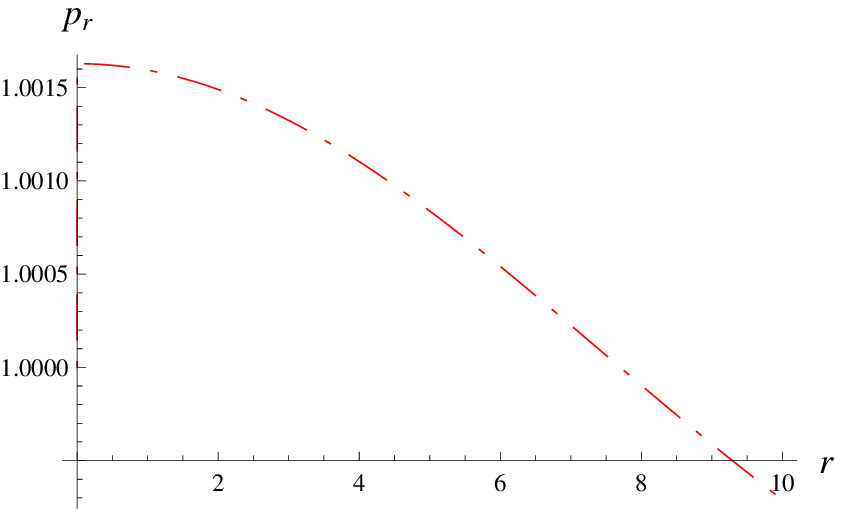, width=0.6\linewidth}
\epsfig{file=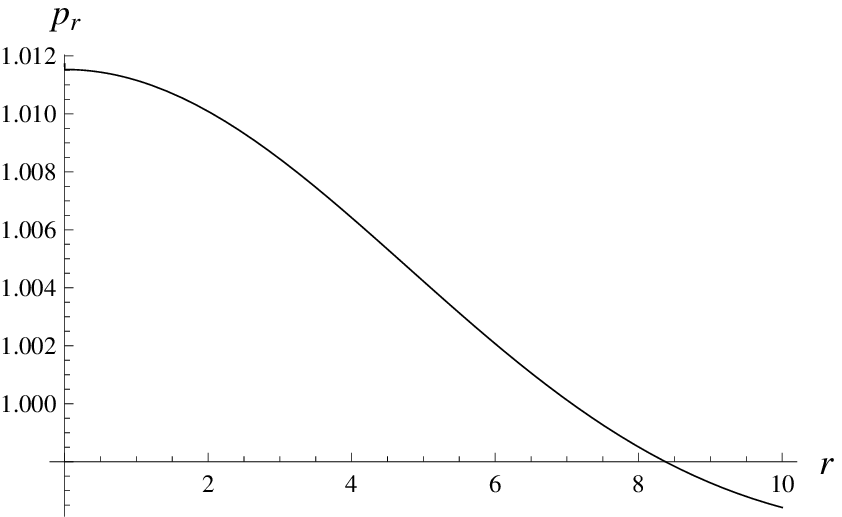, width=0.6\linewidth} \epsfig{file=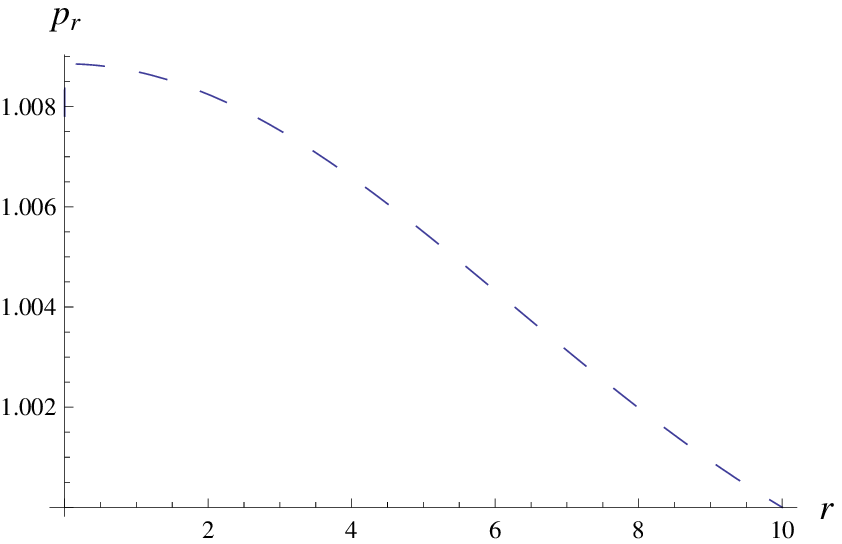,
width=0.6\linewidth}\caption{First, second and third graphs
represent the radial pressure variation of Strange star candidate
Her X-1, SAX J 1808.4-3658(SS1) and 4U 1820 - 30, respectively.}
\end{figure}

\begin{figure}
\center\epsfig{file=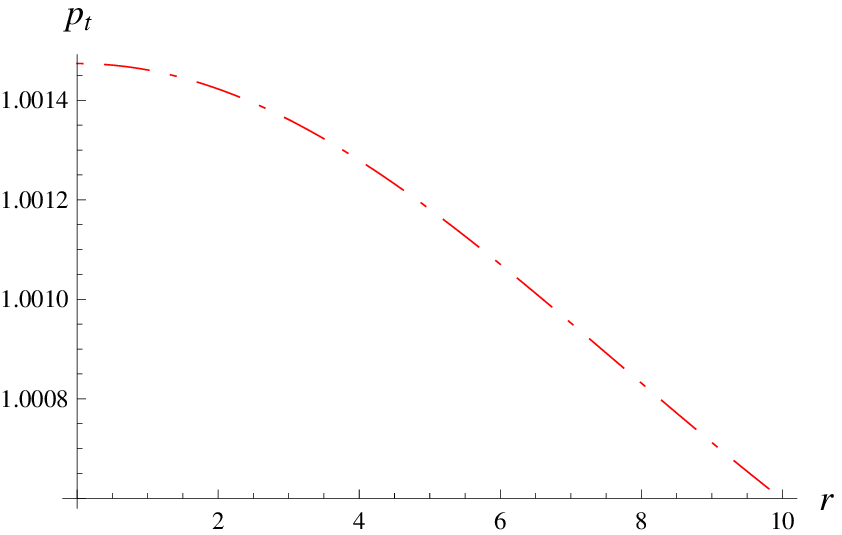, width=0.6\linewidth}
\epsfig{file=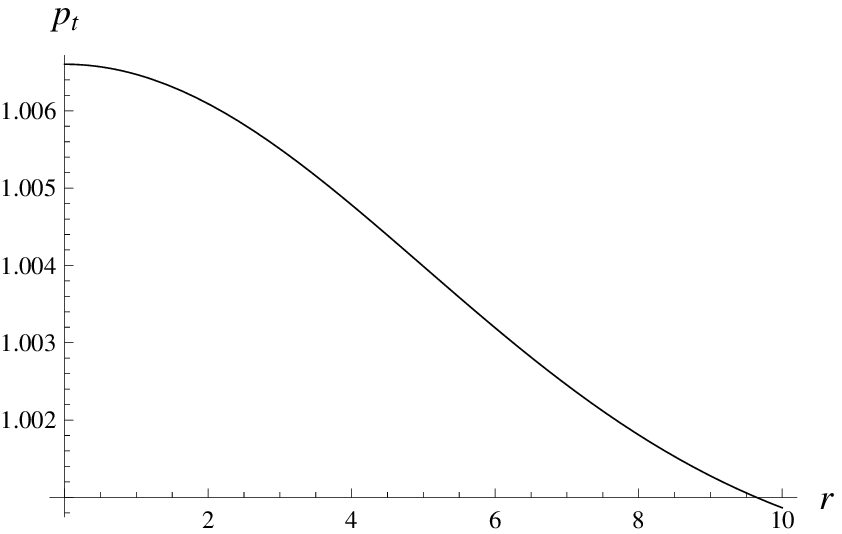, width=0.6\linewidth} \epsfig{file=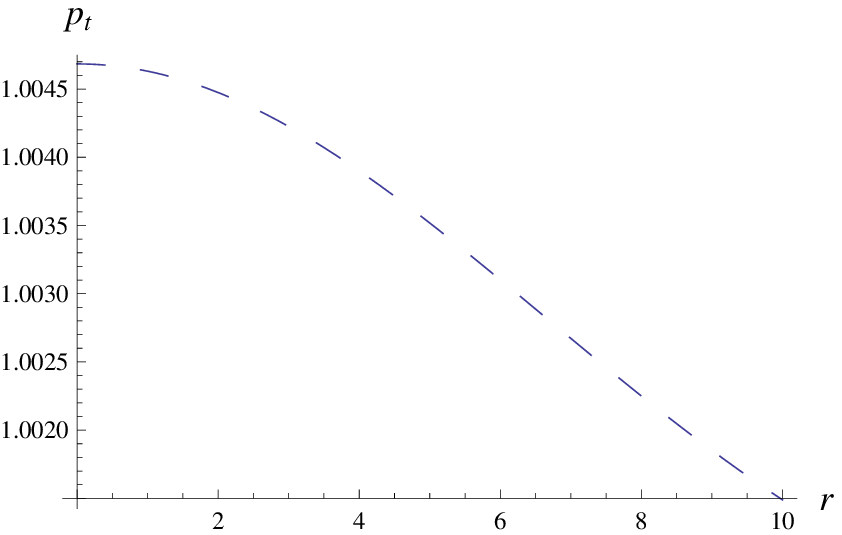,
width=0.6\linewidth}\caption{First, second and third graphs
represent the transverse pressure variation of Strange star
candidate Her X-1, SAX J 1808.4-3658(SS1) and 4U 1820 - 30,
respectively.}
\end{figure}
\begin{figure}
\center\epsfig{file=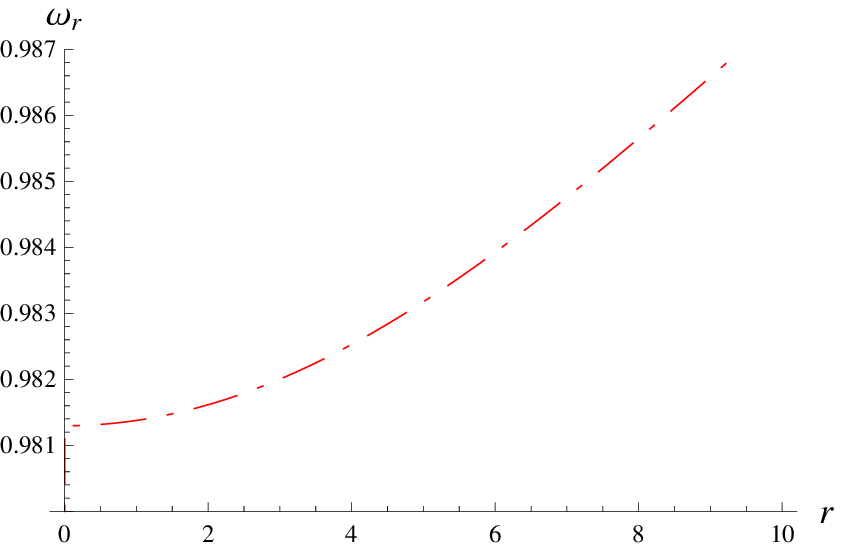, width=0.6\linewidth}
\epsfig{file=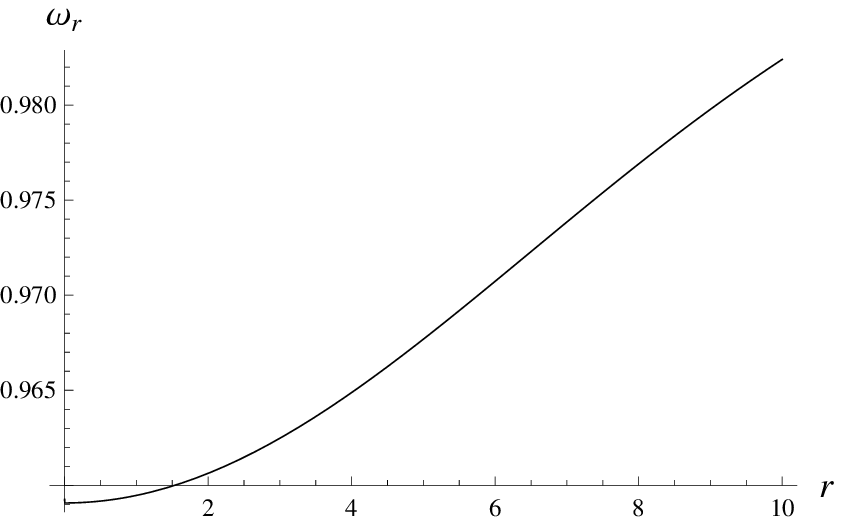, width=0.6\linewidth} \epsfig{file=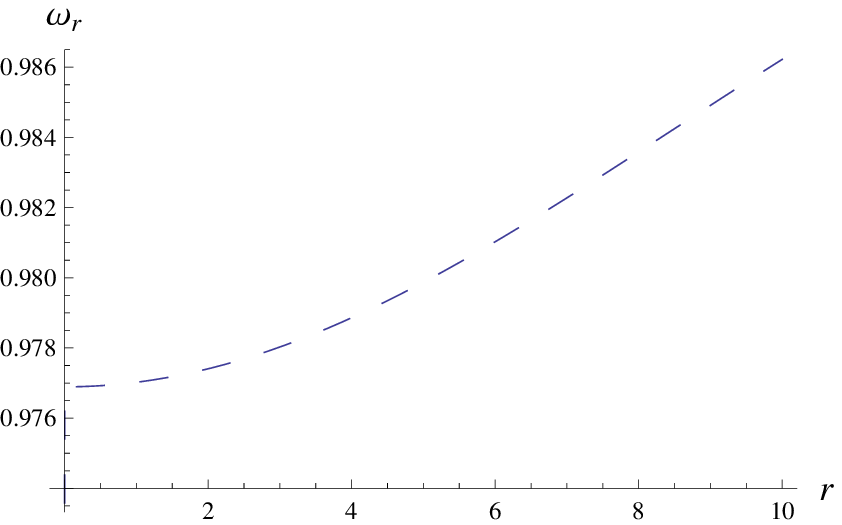,
width=0.6\linewidth}\caption{First, second and third graphs
represent the EOS parameter ${\omega}_{r}$ variation of Strange star
candidate Her X-1, SAX J 1808.4-3658(SS1) and 4U 1820 - 30,
respectively.}
\end{figure}
\begin{figure}
\center\epsfig{file=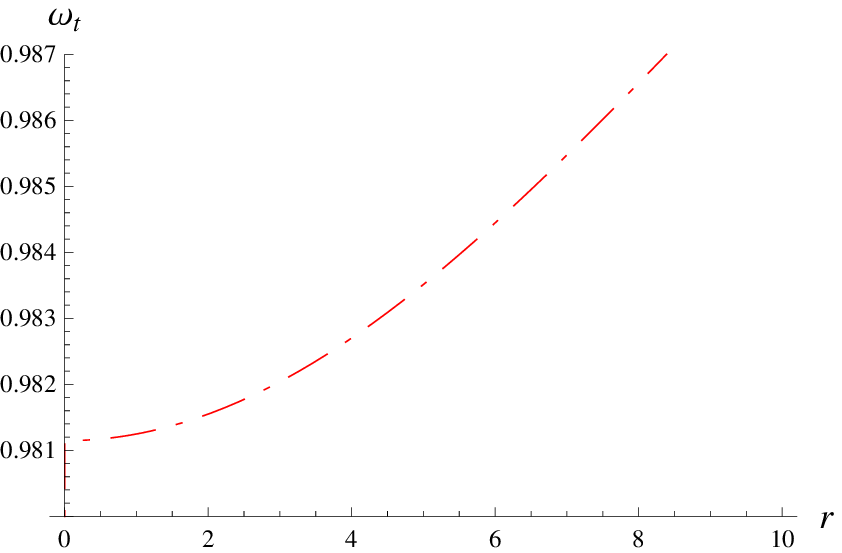, width=0.6\linewidth}
\epsfig{file=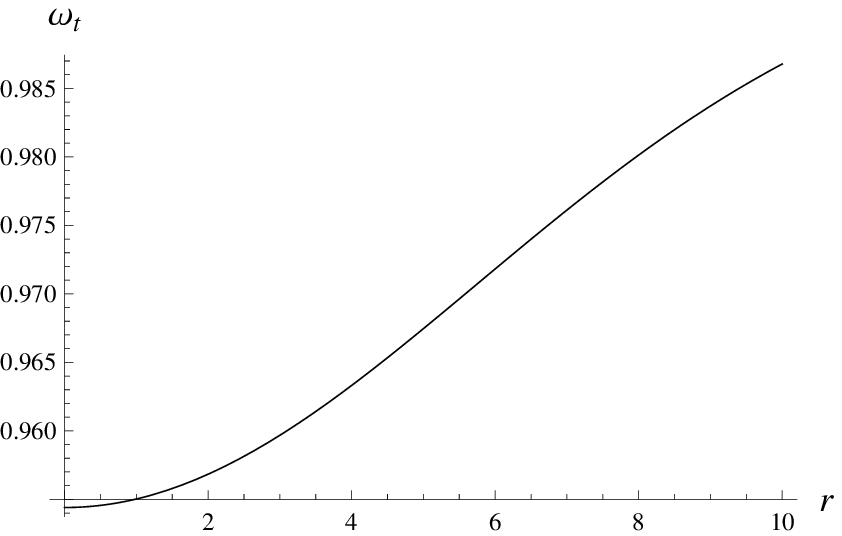, width=0.6\linewidth} \epsfig{file=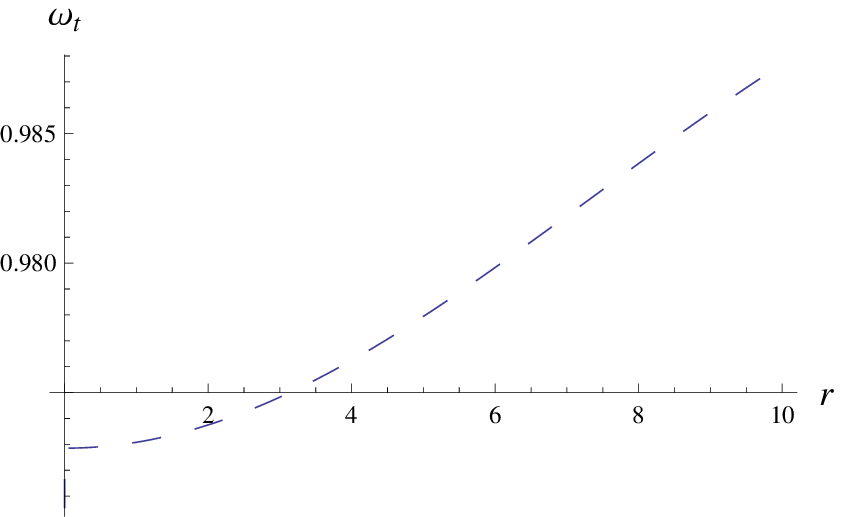,
width=0.6\linewidth}\caption{First, second and third graphs
represent the EOS parameter ${\omega}_{t}$ variation of Strange star
candidate Her X-1, SAX J 1808.4-3658(SS1) and 4U 1820 - 30,
respectively.}
\end{figure}


\section{Physical Analysis}

In this section, we shall discuss following features of our model:

\subsection{Anisotropic Behavior}

From Eqs.(\ref{tt}) and (\ref{rr}), we get the radial gradient of pressures:
\begin{eqnarray}
&&\frac{dp_{r}}{dr}=\frac{1}{k^{2}}\Big[{2e^{-Ar^{2}}\Big(1-e^{Ar^{2}}+Ar^{2}%
\Big(1+(4B+\alpha G^{n}(n-1))\Big)\Big)}\Big]
\end{eqnarray}

\begin{eqnarray}
\frac{d\rho }{dr}&&=\frac{1}{c^{2}k^{2}}\Big(2e^{-3Ar^{2}}\Big(e^{2Ar^{2}}%
\Big(-1+e^{Ar^{2}}+2Ar^{2}\Big)+Ae^{2Ar^{2}}r^{2}\Big(e^{Ar^{2}}2Ar-1\Big)-
\\
&&Ae^{2Ar^{2}}\Big(2+e^{Ar^{2}}\Big) +16A\alpha nG^{n-3}(n-1)r^{2}\Big(%
1+e^{Ar^{2}}\Big(-1+2(B-Br^{2})r^{2}\Big)G^{^{\prime }}  \notag \\
&&(-2+n-\lambda ^{^{\prime }})+G^{^{\prime \prime }})-8\alpha nG^{n-3}(n-1)%
\Big(e^{Ar^{2}}-Ar^{2}-1\Big)\Big(G^{^{\prime }}(n-2-\lambda ^{^{\prime
}})+G^{^{\prime \prime }}\Big)  \notag \\
&&-4\alpha nG^{n-3}(n-1)r\Big(1+e^{Ar^{2}}\Big(-1+2(B-Br^{2})r^{2}\Big)\Big)%
\Big((n-2-\lambda ^{^{\prime }})G^{^{\prime \prime }}\lambda ^{^{\prime
\prime }}+G^{(3)})\Big)\Big)  \notag
\end{eqnarray}

We compute second order derivatives as the following:
\begin{eqnarray}
\frac{d^{2}\rho }{dr^{2}}&&=\frac{1}{c^{2}k^{2}}\Big[ e^{-3Ar^{2}}\Big(%
-2e^{2Ar^{2}}\Big(3(-1+e^{Ar^{2}})+Ar^{2}\Big(-3-2Ar^{2} \\
&&+2Ar^{2}(-1+2Ar^{2})\Big)\Big) -8\alpha nG^{n-3}(n-1)\Big(6G^{^{\prime
\prime }}+G^{^{\prime }}\Big(6(1+3Ar^{2}(1+2Ar^{2})\Big)  \notag \\
&&\Big(-2+n-\lambda ^{^{\prime }}\Big)+4r(1+3Ar^{2})\lambda ^{^{\prime
\prime }}-r^{2}\lambda ^{(3)}+e^{Ar^{2}}(2(n-2)\Big(-3 +2Ar^{2}\Big(-3
\notag \\
&&+2(-B+Br^{2})r^{2}+4Ar^{2}\Big(-1+2(B-Br^{2}) r^{2}\Big)\Big)\Big)+\Big(%
6+4Ar^{2}\Big(3+2(B-Br^{2})r^{2}  \notag \\
&&+4Ar^{2}\Big(1+2 (-B+Br^{2})r^{2}\Big)\Big)\Big)\lambda ^{^{\prime }}+4r%
\Big(-1+2Ar^{2}\Big(-1+2(B-Br^{2})r^{2}\Big)\Big)\lambda ^{\prime \prime }
\notag \\
&& +r^{2}\Big(1+2(-B+Br^{2})r^{2}\Big)\lambda ^{(3)}\Big)\Big)+r\Big(%
2G^{^{\prime \prime }}\Big(4-2n+3Ar \Big(3+4r-2nr  \notag \\
&&+6Ar^{2}\Big)+(2+6Ar^{2})\lambda ^{^{\prime }}-r\lambda ^{^{\prime \prime
}}\Big)-\Big(4+2r-nr +12Ar^{2}+r\lambda ^{^{\prime }}\Big)G^{(3)}+rG^{(4)}%
\Big)  \notag \\
&&+e^{Ar^{2}}\Big(2G^{^{\prime \prime }}\Big(-3+2(n-2)r +2Ar^{2}\Big(-3+2r%
\Big(-2+n+(-B+Br^{2})r  \notag \\
&&-2(B-Br^{2})(n-2)r^{2}\Big) +4Ar^{2}\Big(-1+2(B-Br^{2})r^{2}\Big)\Big)+r%
\Big(\Big(-2+4Ar^{2}\Big(-1  \notag \\
&&+2(B-Br^{2})r^{2}\Big)\lambda ^{^{\prime }} +r\Big(1+2(-B+Br^{2})r^{2}%
\lambda ^{^{\prime \prime }}\Big)\Big)+r\Big(\Big(4+(n-2)r\Big(-1  \notag \\
&&+2(B-Br^{2})r^{2}\Big) +8Ar^{2}\Big(1+2(-B+Br^{2})r^{2})+r\Big(%
1+2(-B+Br^{2})r^{2}\Big)\lambda ^{^{\prime }}\Big)G^{(3)}  \notag \\
&&+r(-1+2 \Big(B-Br^{2}\Big)r^{2}\Big)G^{(4)}\Big)\Big)\Big)\Big) \Big]
\end{eqnarray}

\begin{eqnarray}
&&\frac{d^{2}p_{r}}{dr^{2}}=\frac{1}{k^{2}}\Big[2e^{-Ar^{2}}\Big(3\Big(%
-1+e^{Ar^{2}}\Big)+Ar^{2}\Big(-3+\Big(4B+ \alpha G^{n}(n-1)\Big)r^{2} \\
&&-2Ar^{2}\Big(1+\Big(4B+\alpha G^{n}(n-1)\Big)r^{2}\Big)\Big)\Big)\Big]
\end{eqnarray}

We observe that at center $r=0$, our model provides that
\begin{eqnarray}  \label{se4}
\frac{d\rho}{dr}&=&0,~~~ \frac{dp_r}{dr}=0 \\
\frac{d^2\rho}{dr^2}&<&0,~~~~ \frac{d^2p_r}{dr^2}<0.
\end{eqnarray}
This indicate maximality of radial pressure and density. This fact
implies that $\rho$ and $p_r$ are decreasing function of $r$ as
shown in figures \textbf{1,2} for a class of strange stars. The
measure of anisotropy is
\begin{equation*}
\Delta=\frac{2}{r}(p_t-p_r),
\end{equation*}
which takes the form
\begin{eqnarray}
\Delta &&=\frac{2}{r}(p_t-p_r)\equiv\frac{2}{{\kappa}^2r}\Big[-\frac{%
e^{-Ar^{2}}}{2k^{2}r^{2}}(2r^{2}A+e^{Ar^{2}}-1)+\frac{4e^{-2Ar^{2}}}{k^{2}}%
\Big( \alpha n(n-1)G^{n-3}(G^{^{\prime \prime }}+\lambda ^{^{\prime
}}G^{^{\prime }})  \notag \\
&&+\alpha n(n-1)(n-2)G^{n-3}(G^{^{\prime }})^{2}\Big)\left( \frac{%
1-e^{Ar^{2}}}{r^{2}}-2(B-Br^{2})\right)-3\frac{\alpha }{2k^{2}}G^{n}(n-1)
\notag \\
&&-\frac{e^{-Ar^{2}}}{k^{2}}(B+B^{2}r^{2}-ABr^{2}+\frac{2}{r}(Br-Ar)+\frac{%
1-e^{Ar^{2}}}{r^{2}})-\frac{4e^{-2Ar^{2}}}{k^{2}}\Big(\frac{2Br}{r}  \notag
\\
&&+\frac{1-e^{Ar^{2}}}{r^{2}}\Big(\alpha n(n-1)G^{n-1}(G^{^{\prime \prime
}}-2ArG^{^{\prime }})+\alpha n(n-1)(n-1)G^{n-2}(G^{^{\prime }})^{2}\Big)
\notag \\
&&+\frac{e^{-Ar^{2}}}{2k^{2}}\Big( -\frac{1}{r^{2}} (4r^{2B}-e^{Ar^{2}}+1)+%
\alpha G^{n}(n-1)\Big)-\Big(2\frac{e^{-Ar^{2}}}{2k^{2}}\Big( -\frac{1}{r^{2}}
(4r^{2B}-e^{Ar^{2}}+1)  \notag \\
&&+\alpha G^{n}(n-1)\Big)\Big)\Big]
\end{eqnarray}

The anisotropy will be directed outward when $p_t>p_r$ this implies that $%
\Delta>0$ and directed inward when $p_t<p_r$ implying $\Delta>0$. In
this case $\Delta>0$, for larger values of $r$ for a class of
strange stars as shown in figures \textbf{6}. This implies that
anisotropic force allows the construction of more massive stars. In
order to comprehend some general results associated with the strong
gravitational fields, we include weak energy condition (WEC), null
energy condition (NEC), strong energy condition (SEC) and dominant
energy condition (DEC). For an anisotropic fluid, these are defined
as
\begin{eqnarray}
\mathbf{NEC}:\quad&&\rho+p_r\geq0, \quad \rho+p_t\geq0,  \notag \\
\mathbf{WEC}:\quad&&\rho\geq0, \quad \rho+p_r\geq0, \quad \rho+p_t\geq0,
\notag \\
\mathbf{SEC}:\quad&&\rho+p_r\geq0, \quad \rho+p_t\geq0, \quad
\rho+p_r+2p_t\geq0,  \notag \\
\mathbf{DEC}:\quad&&\rho>|p_r|, \quad \rho>|p_t|.  \notag
\end{eqnarray}
We find that our model satisfies these conditions (\textbf{see
Figs.6-8})for specific values of mass and radius which helps to find
the unknown parameters for different strange stars.

\begin{figure}
\centering \epsfig{file=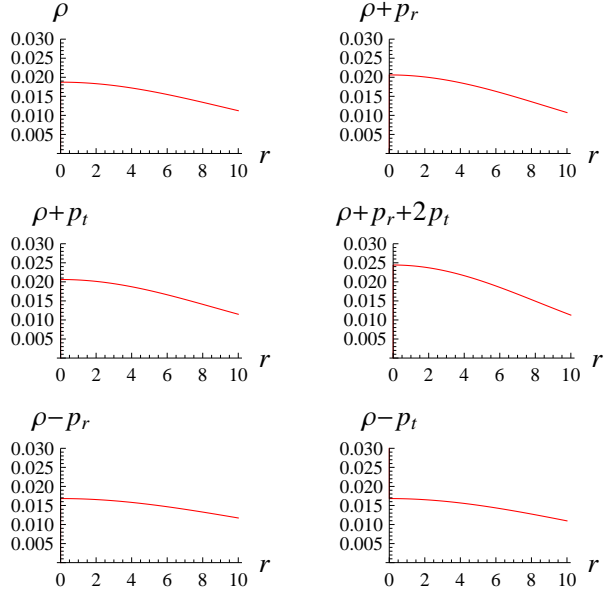}\caption{Evolution of energy
constraints at the stellar interior of strange star Her X-1.}
\end{figure}

\begin{figure}
\centering \epsfig{file=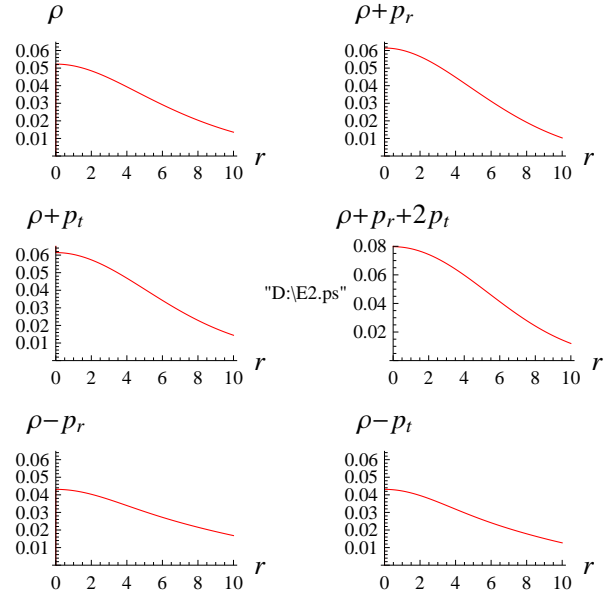}\caption{Evolution of energy
constraints at the stellar interior of strange star SAX J
1808.4-3658.}
\end{figure}

\begin{figure}
\centering \epsfig{file=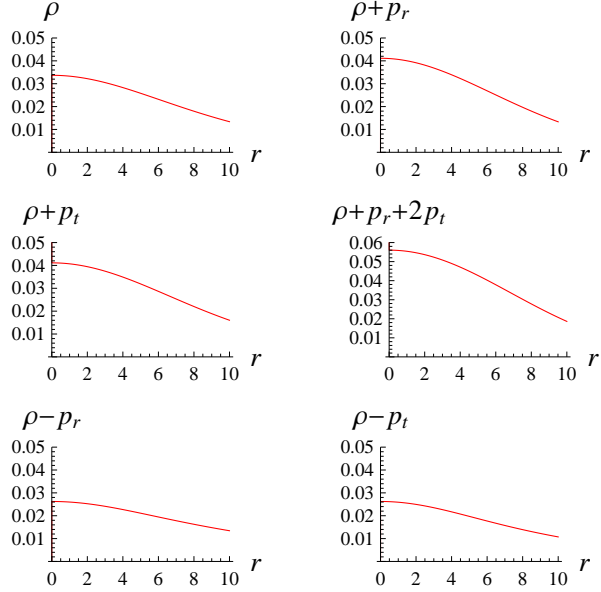}\caption{Evolution of energy
constraints at the stellar interior of strange star 4U 1820-30.}
\end{figure}



\subsection{Matching Conditions}


The intrinsic metric of the boundary surface will be the same
whether it is derived from the internal or external geometry of the
star. It confirms that for any coordinate system the metric tensor
components will be continuous across the boundary surface. for this
purpose the matching conditions are required for the interior metric
(\ref{g}), so that the system of field equations can be solved
constrained to the boundary conditions that radial pressure $p_r =
0$ at $r = R$. The interior metric (\ref{g}) can be matched smoothly
at $r = R$, to an exterior Schwarzschild metric given by
\begin{equation}  \label{21}
ds^2=\left(1-\frac{2M}{r}\right)dt^2-\left(1- \frac{2M}{r}%
\right)^{-1}dr^2-r^2(d\theta^2+\sin^2\theta d\varphi^2).
\end{equation}
At the boundary $r=R$ continuity of the metric functions $g_{tt}$, $g_{rr}$
and $\frac{\partial g_{tt}}{\partial r}$ at the boundary surface yield,
\begin{eqnarray}  \label{22}
g_{tt}^-=g_{tt}^+,~~~~~ g_{rr}^-=g_{rr}^+,~~~~~ \frac{\partial g_{tt}^-}{%
\partial r}=\frac{\partial g_{tt}^+}{\partial r},
\end{eqnarray}
where $-$ and $+$, correspond to interior and exterior solutions. From the
interior and exterior matrices, we get
\begin{eqnarray}  \label{24}
A&=&-\frac{1}{R^2}ln\left(1-\frac{2M}{r}\right), \\
B&=&\frac{M}{R^3}{{\left(1-\frac{2M}{r}\right)}^{-1}} \\
C&=&ln\left(1-\frac{2M}{r}\right)-\frac{M}{R}{{\left(1-\frac{2M}{r}\right)}%
^{-1}}
\end{eqnarray}
For the given values of $M$ and $R$ for given star, the constants
$A$ and $B$ are given in the table \textbf{1}. We would like to
mention that the values of $M$ and $R$ have been taken from Lattimer
et al. (2014), Li et al. (1999), Leahy et al. (2007) to calculate
the values of $A$ and $B$.

\begin{table}[ht]
\caption{Values of constants $A$ and $B$ for the given values of $M$
and $R$ of compact star}
\begin{center}
\begin{tabular}{|c|c|c|c|c|c|}
\hline {Strange Quark Star}&  \textbf{ $M$} & \textbf{$R(km)$} &
\textbf{ $\frac{M}{R}$} &\textbf{ $A(km ^{-2})$}& \textbf{$B(km
^{-2})$}
\\\hline  Her X-1& 0.88$M_\odot$& 7.7&0.168&0.006906276428 &
$0.004267364618$
\\\hline SAX J 1808.4-3658& 1.435$M_\odot$& 7.07&0.299& 0.01823156974 &
$0.01488011569$
\\\hline 4U 1820-30&2.25$M_\odot$& 10.0 &0.332&0.01090644119 &
$0.009880952381$
\\\hline
\end{tabular}
\end{center}
\end{table}

\subsection{Stability}

Here, we discuss the stability of strange stars for this purpose the sound
speeds $V_{rv}^2$ and $V_{tv}^2$ along along radial and transverse
directions are given by
\begin{eqnarray}
V_{rv}^{2}&&=\frac{dp_{r}}{d\rho }\equiv \Big[\frac{1}{k^{2}r^{3}}\Big[{%
2e^{-Ar^{2}}\Big(1-e^{Ar^{2}}+Ar^{2}\Big(1+(4B+\alpha G^{n}(n-1))\Big)\Big)}%
\Big]\times\Big[\frac{1}{c^{2}k^{2}r^{3}}  \notag \\
&&\times \Big(2e^{-3Ar^{2}}\Big(e^{2Ar^{2}}\Big(-1+e^{Ar^{2}}+2Ar^{2}\Big)%
+Ae^{2Ar^{2}}r^{2}\Big(e^{Ar^{2}}2Ar-1\Big)  \notag \\
&&-Ae^{2Ar^{2}}\Big(2+e^{Ar^{2}}\Big) +16A\alpha nG^{n-3}(n-1)r^{2}\Big(%
1+e^{Ar^{2}}\Big(-1+2(B-Br^{2})r^{2}\Big)G^{^{\prime }}  \notag \\
&&(-2+n-A)+G^{^{\prime \prime }})-8\alpha nG^{n-3}(n-1)\Big(%
e^{Ar^{2}}-Ar^{2}-1\Big)\Big(G^{^{\prime }}(n-2-A)+G^{^{\prime \prime }}\Big)
\notag \\
&&-4\alpha nG^{n-3}(n-1)r\Big(1+e^{Ar^{2}}\Big(-1+2(B-Br^{2})r^{2}\Big)\Big)%
\Big((n-2-A)G^{^{\prime \prime }}A+G^{(3)})\Big)\Big)\Big]^{-1},  \notag \\
\\
V_{tv}^{2} &=&\frac{dp_{t}}{d\rho }\equiv \Big[\frac{1}{k^{2}r^{3}}\Big[{%
2e^{-Ar^{2}}\Big(1-e^{Ar^{2}}+Ar^{2}\Big(1+(4B+\alpha G^{n}(n-1))\Big)\Big)}%
\Big]\Big]\times\Big[\frac{1}{c^{2}k^{2}r^{3}}  \notag \\
&&\times \Big(2e^{-3Ar^{2}}\Big(e^{2Ar^{2}}\Big(-1+e^{Ar^{2}}+2Ar^{2}\Big)%
+Ae^{2Ar^{2}}r^{2}\Big(e^{Ar^{2}}2Ar-1\Big)  \notag \\
&&-Ae^{2Ar^{2}}\Big(2+e^{Ar^{2}}\Big) +16A\alpha nG^{n-3}(n-1)r^{2}\Big(%
1+e^{Ar^{2}}\Big(-1+2(B-Br^{2})r^{2}\Big)G^{^{\prime }}  \notag \\
&&(-2+n-\lambda ^{^{\prime }})+G^{^{\prime \prime }})-8\alpha nG^{n-3}(n-1)%
\Big(e^{Ar^{2}}-Ar^{2}-1\Big)\Big(G^{^{\prime }}(n-2-A)+G^{^{\prime \prime }}%
\Big)  \notag \\
&&-4\alpha nG^{n-3}(n-1)r\Big(1+e^{Ar^{2}}\Big(-1+2(B-Br^{2})r^{2}\Big)\Big)%
\Big((n-2-A^2)G^{^{\prime \prime }}A+G^{(3)})\Big)\Big)\Big]^{-1}.  \notag
\end{eqnarray}

From above equations, we have

\begin{eqnarray}
V_{tv}^{2}-V_{rv}^{2}&&=\Big[\lbrack\ \Big(c^{2}e^{2Ar^{2}}\Big(-2\Big(%
1-e^{Ar^{2}}\Big(1+\Big(4B+\alpha G^{n}\Big)(n-1)\Big)r^{2}\Big)\Big)%
+e^{Ar^{2}}  \notag \\
&&\times \Big(-1+e^{-Ar^{2}}-2Ae^{-Ar^{2}}r^{2}+e^{-Ar^{2}}\Big(%
-1+e^{Ar^{2}}+2Ar^{2}\Big)+Ae^{-Ar^{2}}r^{2}  \notag \\
&& \Big(-1+e^{Ar^{2}}+2Ar^{2}\Big)-2Ae^{-2Ar^{2}}r^{2}\Big(-1+e^{Ar^{2}}\Big(%
1+r\Big(-2Ar+2Br-ABr^{3}  \notag \\
&& Br+B^{2}r^{3}\Big)\Big)\Big))-2e^{-2Ar^{2}}\Big(1+e^{Ar^{2}}\Big(%
-1+Ar^{2}-Br^{2}+B^{2}r^{4}\Big)+Ar^{2}\Big)  \notag \\
&& -8\alpha ne^{-2Ar^{2}}G^{n-3}(n-1)\Big(1+e^{Ar^{2}}(-1+Ar^{2})\Big)\Big(%
(n-2)\Big(G^{^{\prime }}\Big)^{2}-G^{^{\prime }}A-G^{^{\prime \prime }}\Big)
\notag \\
&& +16A\alpha ne^{-3Ar^{2}}G^{n-2}(n-1)r^{2}\Big(-1+e^{Ar^{2}}(1+2Br^{2})%
\Big)\Big(2GarG^{^{\prime }}-(n-1)(G^{^{\prime }})^{2}  \notag \\
&& -GG^{^{\prime \prime }}\Big)+\Big[2\alpha ne^{-2Ar^{2}}G^{n-3}(n-1)r\Big(%
\alpha G^{n}(n-1)r^{2}-2k^{2}\Big(-1+e^{Ar^{2}}  \notag \\
&&+2-GG^{^{\prime \prime }}\Big)\Big)\Big] +\Big[(B-Br^{2})r^{2}\Big(%
-AG^{^{\prime \prime }}+G^{^{\prime }}\Big(2(n-2)G^{\prime \prime }-A\Big)%
G^{(3)}\Big) \Big]  \notag \\
&&\times\Big[4\alpha ne^{-3Ar^{2}}G^{n-2}(n-1)r\Big(-1+e^{Ar^{2}}(1+2Br^{2})%
\Big)\Big(G\Big(G^{(3)}-2ArG^{^{ \prime }}\Big)  \notag \\
&& +2G^{^{\prime }}\Big((n-1)G^{^{\prime \prime }}-GArG^{^{\prime }}\Big)%
\Big)\Big]^{-1}\Big]  \notag \\
&&\times\Big[%
(2(e^{2Ar^{2}}(-1+e^{Ar^{2}}+2Ar^{2})+Ae^{2Ar^{2}}r^{2}(-1+e^{Ar^{2}}+2Ar^{2})-Ae^{2Ar^{2}}r^{2}
\notag \\
&& \Big(2+e^{Ar^{2}}\Big)+16A\alpha nG^{n-3}(n-1)r^{2}\Big(%
1+e^{Ar^{2}}(-1+(B-Br^{2})r^{2}\Big)\Big)  \notag \\
&& \Big(G^{^{\prime }}(-2+n-A)+G^{^{\prime \prime }}\Big)-8\alpha
nG^{n-3}(n-1)\Big(-1+e^{Ar^{2}}-Ar^{2}\Big)  \notag \\
&& \Big(G^{^{\prime }}(-2+n-A)G^{^{\prime \prime }}\Big)-4\alpha
nG^{n-3}(n-1)r\Big(1+e^{Ar^{2}}\Big(-1+2  \notag \\
&& (B-Br^{2})r^{2}\Big)\Big)\Big(\Big(-2+n-A)G^{^{\prime \prime
}}-G^{^{\prime }}A+G^{(3)}\Big)\Big)\Big)\Big]^{-1}.  \notag
\end{eqnarray}
From figure \textbf{6}, we can see that $\mid\upsilon^2_{st}-%
\upsilon^2_{sr}\mid\leq1$. This is used to check whether local
anisotropic matter distribution is stable or not. For purpose, we
use the cracking concept introduced by Herrera (1992) which can
examine that potentially stable region is such region where radial
speed of sound is greater than the transverse speed of sound. Hence,
our proposed compact star model is stable.


\begin{figure}
\center\epsfig{file=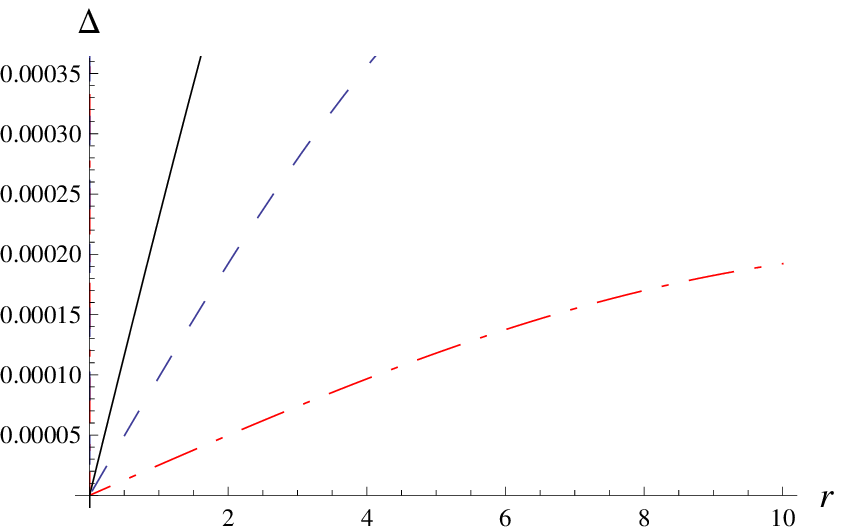, width=0.6\linewidth}
\epsfig{file=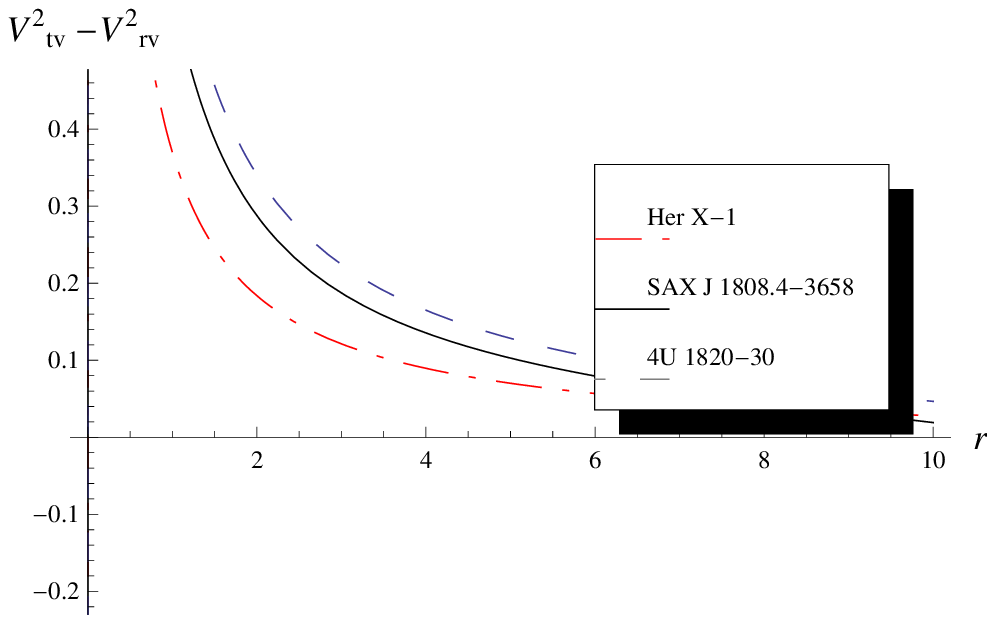, width=0.6\linewidth} \caption{First and second
graphs represent the variation of $\Delta$ and $V^2_{tv}-V^2_{rv}$,
respectively.}
\end{figure}


\section{Conclusion}

On the basis of cosmological observations it has been concluded that
that there are two phases of accelerated expansion in our universe:
cosmic inflation in early universe and acceleration in the current
expansion of the Universe. The investigation of current cosmic
expansion and nature of DE has been widespread among the scientists.
For this purpose, numerous efforts have been made based upon
different strategies to modify the General Relativity. The $f(G)$
gravity is one of the modifications of General Relativity.

The present paper deals with the formulation of analytical models of compact
stars with the anisotropic gravitating static source in the framework of $%
f(G)$ gravity. For this purpose, we have considered the power law form of $%
f(G)$ gravity model, further the stars are assumed as anisotropic in their
internal structure. The analytic solution in $f(G)$ gravity have found by
matching the interior metric with the well-known exterior metric. The
physical analysis of the results implies the following properties of the
anisotropic compact stars in $f(G)$ gravity:

\begin{itemize}
\item The bound on the EoS parameters are given by $0<{\omega}_t<1$ and $0<{\omega}_r<1$, which is
consistent with normal matter distribution in $f(G)$ gravity.

\item The density and pressures attain the maximum value at the center.
Hence, these are decreasing functions.

\item It has been found that the anisotropy will be directed outward when $%
p_t>p_r$ this implies that $\Delta>0$ and directed inward when
$P_t<P_r$ implying $\Delta<0$. In this case $\Delta>0$, for larger
values of $r$ for various strange stars. This implies that
anisotropic force favors the construction of more massive star in
$f(G)$ gravity.

\item The variation of $V^2_{tv}-V^2_{rv}$ of different strange stars is
shown in Figure \textbf{6}. This implies that difference of the two
sound
speeds, i.e., $V^2_{tv}-V^2_{rv}$ satisfies the inequality $%
|V^2_{tv}-V^2_{rv}|\leq1$. Thus, our proposed strange star model is stable
in $f(G)$ gravity.
\end{itemize}


\newpage

\title{\bf Response of the Referee's Report about the paper entitled\\
Anisotropic Compact Stars in $f(G)$ Gravity}

Authors: G. Abbas, D. Momeni, M. Aamir Ali,  R. Myrzakulov, and
S.Qaiasr

\maketitle

We have inserted the relevant explanation in the revised version in
the light of referee's comments. These are given in the following:\\\\
({\textbf{Response-Major Corrections}})\\
\par \noindent
\begin{enumerate}
\item All the indicated and many other typos have been removed
\item Here, we have provided the references i.e., and later revisited by Jedamzik
et al. (2010) and Martin et al. (2014).
\item Here we have also provided  reference i.e., It has been demonstrated (Capozziello and De Laurentis
2011, Nojiri and Odintsov 2011)
\item For the discussion of Anisotropic compact stars we have
presented the following paragraph at page \textbf{3} and have
mentioned the required reference

"A class of exact solutions of Einstein field equations with
anisotropic source was presented by Mak and Harko (2004). These
solutions show that energy density as well as tangential and radial
pressure are finite and positive inside the anisotropic star. Chaisi
and Maharaj (2005) formulated an algorithm for the solutions of the
field equations with anisotropic matter source. By using Chaplygin
gas equation of state (EOS), Rahaman et al. (2012) extended the
Krori-Barua (1975) solution to the charge anisotropic source.
Recently (Kalam et al. 2012, Kalam et al. 2013) there has been
growing interest to study the models of the compact objects using
the anisotropic source with Krori-Barua metric. Hossein et al.
(2012) formulated the anisotropic star model with varying
cosmological constant. Bhar et al. (2015) discuss the possibility
for the existence of higher dimensional compact star. Maurya et
al.(2015) formulated the charged anisotropic solutions for the
compact stars."

\item  For the Krori-Barua ansatz at page \textbf{5}, we have added
the following motivation

"Indeed this solution was treated as a singularity free exact
solution for astatic charged fluid sphere. So our main motivation to
introduce this set of functions for static metric goes back to the
singularity free structure of our extended model for compact star.
It is easy to show that neither densities nor curvature terms become
singular in the vicinity of $r=0$ with this appropriate set of
metric functions."

also, for the ansatz  given by Eq.(27) at page  13, we have added

The intrinsic metric of the boundary surface will be the same
whether it is derived from the internal or external geometry of the
star. It confirms that for any coordinate system the metric tensor
components will be continuous across the boundary surface. for this
purpose the matching conditions are required for the interior metric
(\ref{g}), so that the system of field equations can be solved
constrained to the boundary conditions that radial pressure $p_r =
0$ at $r = R$. The interior metric (\ref{g}) can be matched smoothly
at $r = R$, to an exterior Schwarzschild metric.
\item All the plots in the Figs. 1, 2, 3, 4, 5 and 9 have been
enlarged for clarity
\item Energy conditions for the different compact stars have been
shown in  Figs. 6-8.
\item We have modified the table 1 by adding few column, data set from Lattimer et al.
(2014) have been used for masses and radii  of the  stars. The
values of the constants in table have been calculated from our
models. The references are only for observed values of mass and
radius of each star. To, avoid any confusion it all has been
clarified before table \textbf{1}.
\end{enumerate}

Finally, we would like to thank the Referee for his keen interest to
study the paper in detail. Also, his fruitful comments are highly
acknowledged.

\begin{thebibliography}{99}
\bibitem{*}

Abbas. G. , Nazeer. S., Meraj. M. A.: Astrophysics and Space Science
\textbf{345}, 449(2014)

Arapoglu. A.~S.~, Deliduman. C. , Eksi. K.~Y.~,
JCAP \textbf{1107}, 020 (2011) [arXiv:1003.3179 [gr-qc]].

Astashenok. A.~V.~, Capozziello. S. , Odintsov. S.~D.~,
arXiv:1401.4546 [gr-qc]. 

Astashenok. A.~V.~, Capozziello. S.~, Odintsov. S.~D.~,
JCAP \textbf{1312}, 040 (2013) [arXiv:1309.1978 [gr-qc]].

Astashenok. A.~V.~, Capozziello. S.~, Odintsov. S.~D.~,
arXiv:1405.6663 [gr-qc]. 

Astashenok. A.~V., Capozziello. S.~, Odintsov. S.~D.~,
arXiv:1408.3856 [gr-qc].

Bamba. K.~, Nojiri. S.~'i.~, Odintsov. S.~D.~,
arXiv:1302.4831 [gr-qc]. 

Bamba. K.~ and Odintsov. S.~D.~,
arXiv:1402.7114 [hep-th].

Bamba. K.~, et al.,
Eur.\ Phys.\ J.\ C \textbf{67}, 295 (2010) [arXiv:0911.4390 [hep-th]].

Bamba. K.~, et al.,
Europhys.\ Lett.\ \textbf{89}, 50003 (2010) [arXiv:0909.4397 [hep-th]].

Bamba. K., Nojiri. S.~, Odintsov. S.~D.~,
JCAP \textbf{0810} (2008) 045 [arXiv:0807.2575 [hep-th]].

Buchdahl. H. A., Mon. Not. Roy. Astron. Soc. 150, 1 (1970).

Bhar, P., Rahaman, F., Ray, S., Chatterjee, V.: arXiv:1503.03439

Camenzind. M. , "Compact Objects in Astrophysics", Springer-Verlag Berlin
Heidelberg (2007).

Capozziello. S.~ and Laurentis. M.~De , 
Phys.\ Rept.\ \textbf{509}, 167 (2011) [arXiv:1108.6266 [gr-qc]].

Capozziello. S.~et al.,
arXiv:1110.5026 [astro-ph.CO]. 

Capozziello. S. and Laurentis. M.~De ,
Found.\ Phys.\ \textbf{40} (2010) 867 [arXiv:0910.2881 [hep-th]].

Capozziello. S. and Laurentis. M.~De ,
Found.\ Phys.\ \textbf{40} (2010) 867 [arXiv:0910.2881 [hep-th]].

Capozziello. S.~ and Francaviglia. M.~,
Gen.\ Rel.\ Grav.\ \textbf{40} (2008) 357 [arXiv:0706.114

Capozziello. S, Lobo. F.~S.~N.~ and Mimoso. J.~P.~,
arXiv:1407.7293 [gr-qc]. 

Capozziello, S., Laurentis, M.~De and Odintsov. S.~D.~,
arXiv:1406.5652 [gr-qc]. 

Capozziello, S., De Laurentis, M.: Phys. Report \textbf{509},
167(2011)

Capozziello. S, Francaviglia. M.~ and Makarenko. A.~N.~,
Astrophys.\ Space Sci.\ \textbf{349} (2014) 603 [arXiv:1304.5440 [gr-qc]].

Capozziello. S.~, Makarenko. A.~N.~ and Odintsov. S.~D.~,
Phys.\ Rev.\ D \textbf{87} (2013) 8, 084037 [arXiv:1302.0093 [gr-qc]].

Capozziello. S.~, Elizalde. E.~, Nojiri . S.~and Odintsov. S.~D.~,
Phys.\ Lett.\ B \textbf{671} (2009) 193 [arXiv:0809.1535 [hep-th]].

Cognola. G.~, et al.,
Eur.\ Phys.\ J.\ C \textbf{64}, 483 (2009) [arXiv:0905.0543 [gr-qc]].

Cognola. G.~, et al.,
Phys.\ Rev.\ D \textbf{75} (2007) 086002 [hep-th/0611198].

Cognola. G.,~et al.,
Phys.\ Rev.\ D \textbf{73}, 084007 (2006) [hep-th/0601008].

Cooney. A.~, DeDeo. S., Psaltis. D.~,
Phys.\ Rev.\ D \textbf{82}, 064033 (2010) [arXiv:0910.5480 [astro-ph.HE]].

Cvetic. M.~, Nojiri. S.~ and Odintsov. S.~D.~,
Nucl.\ Phys.\ B \textbf{628} (2002) 295 [hep-th/0112045].

Chaisi, M., Maharaj S.D.: General Relativ. Gravit. \textbf{37},
1177(2005)

Deliduman. C.~, Eksi. K.~Y.~ and Keles. V.~,
JCAP \textbf{1205}, 036 (2012) [arXiv:1112.4154 [gr-qc]].

Harko. T.~ and Lobo. F.~S.~N.~, 
Eur.\ Phys.\ J.\ C \textbf{70}, 373 (2010) [arXiv:1008.4193 [gr-qc]].


Harko. T.~,et al., 
Phys.\ Rev.\ D \textbf{84}, 024020 (2011) [arXiv:1104.2669 [gr-qc]].


Herrera. L. , Phys. Lett. \textbf{A165}, 206(1992)

Horndeski. G. W. , Int. J. Theor. Phys. 10, 363-384 (1974).

Houndjo. M.~J.~S.~et al.,%
arXiv:1301.4642 [gr-qc].

Hossein, Sk.,M., Rahaman, F., Naskar, J., Kalam, M., Ray, S.: Int.
J. Mod. Phys. \textbf{D21}, 1250088(2012)

Jedamzik, K., Lemoine, M., Martein, J.: JCAP \textbf{1009}, 038(201)

Krori, K.D. and Barua. J. : J. Phys. A.: Math. Gen.\textbf{\ 8},
508(1975)

Kalam, M., Rahaman, F., Ray, S., Hossein, Sk. M., Karar, I., Naskar,
J.: Eur. Phys. J. \textbf{C72}, 2248(2012)


Kalam, M., Rahaman, F.,  Hossein, Sk. M., Ray, S.: Eur. Phys. J. C,
73, 2409(2013)


Lattimer . J.~M.~and Steiner. A.~W.~,
  Astrophys.\ J.\  {\bf 784}, 123 (2014)
  [arXiv:1305.3242 [astro-ph.HE]].

Li, X.~D.~, Bombaci. I.~, Dey. M.~, Dey. J.~ and van den Heuvel.
E.~P.~J.~,
  Phys.\ Rev.\ Lett.\  {\bf 83}, 3776 (1999)
  [hep-ph/9905356].

Lidsey, J.~E.~, Nojiri . S.~and ~Odintsov. S.~D.,
JHEP \textbf{0206} (2002) 026 [hep-th/0202198]. 

Leahy. D.~A.~, Morsink. S.~M.~ and Cadeau. C.~,
  Astrophys.\ J.\  {\bf 672}, 1119 (2008)
  [astro-ph/0703287 [ASTRO-PH]].

Momeni, D.~ and Myrzakulov. R.~,
arXiv:1408.3626 [gr-qc].

Mak, M.K., Harko, T.: Int. J. Mod. Phys. \textbf{D13}, 149(2004)

 Myrzakulov, R. , Sebastiani. L., Zerbini. S.
, Gen. Rel. Grav. 45, 675 (2013) [arXiv:1208.3392 [gr-qc]];

Martin, J., Ringeval, C., Vennin, V.: JCAP 1410, 038(2014)

Maurya, S. K., Gupta, Y.K., Ray, S.: arXiv: 1502.01915

Nojiri, S.~and S.~D.~Odintsov,
Phys.\ Rept.\ \textbf{505}, 59 (2011) [arXiv:1011.0544 [gr-qc]].

Nojiri. S.~'i.~ and Odintsov. S.~D,
AIP Conf.\ Proc.\ \textbf{1115}, 212 (2009) [arXiv:0810.1557 [hep-th]].

Nojiri, S.~ and Odintsov. S.~D.~, eConf C 0602061, 06 (2006) [Int.
J. Geom. Meth. Mod. Phys. 4, 115 (2007)] [hep-th/0601213].

Nojiri, S.~, Odintsov. S.~D.~ and Tretyakov. P.~V.~,
Phys.\ Lett.\ B \textbf{651} (2007) 224 [arXiv:0704.2520 [hep-th]].

Nojiri. S.~ and Odintsov. S.~D.~,
J.\ Phys.\ Conf.\ Ser.\ \textbf{66} (2007) 012005 [hep-th/0611071].

Nojiri, S.~, Odintsov. S.~D.~ and Sasaki. M.~,
Phys.\ Rev.\ D \textbf{71} (2005) 123509 [hep-th/0504052].

Nojiri. S.~, Odintsov S.~D.~ and Gorbunova. O.~G.~,
J.\ Phys.\ A \textbf{39} (2006) 6627 [hep-th/0510183].

Nojiri. S.~, Odintsov. S.~D.~, Sami. M.~,
Phys.\ Rev.\ D \textbf{74} (2006) 046004 [hep-th/0605039].

Nojiri. S. , Odintsov. S. D. , Phys.Lett. B631,1(2005). 

Nojiri, S.~, 
Grav.\ Cosmol.\ \textbf{9}, 71 (2003) [hep-th/0210056].

Nojiri, S.~, Odintsov. S.~D. , Ogushi. S.~,
Int.\ J.\ Mod.\ Phys.\ A \textbf{17}, 4809 (2002) [hep-th/0205187].

Nojiri. S. , Odintsov. S. D.: Phys. Report \textbf{505}, 59(2011).


Perlmutter, S. et al., Nature 391, 51 (1998).

Riess, A.G. et al., Astron. J. 116, 1009 (1998).

Riess, A.G. et al., Astrophys. J. 536, 62 (2000).

Rahaman, F., Sharma, Ray, S., Maulick, R., Karar, I.: Eur. Phys. J.
\textbf{C72}, 2071(2012)

Rodrigues, M.~E.~et al.,
Can.\ J.\ Phys.\ \textbf{92}, 173 (2014) [arXiv:1212.4488 [gr-qc]].


Sebastiani. L.~, et al.,
Phys.\ Rev.\ D \textbf{88}, no. 10, 104022 (2013) [arXiv:1305.4231 [gr-qc]].

Sharif. M. and Zubair. M. : JCAP \textbf{03}(2012)028; J. Exp. Theor. Phys.
\textbf{117}(2013)248; J. Phys. Soc. Jpn. \textbf{81}(2012)114005; ibid.
\textbf{82}(2013)014002; \textbf{82}(2013)064001; Astrophys. Space Sci.
\textbf{349}(2014)529; Gen. Relativ. Gravit. \textbf{46}(2014)1723.

Sharif. M.~ and Abbas. G.~,
Eur.\ Phys.\ J.\ Plus \textbf{128}, 102 (2013) [arXiv:1308.5675 [gr-qc]].

Sharif. M.~ and Abbas. G.~,
J.\ Phys.\ Soc.\ Jap.\ \textbf{82}, 034006 (2013) [arXiv:1302.1173 [gr-qc]].



\end{thebibliography}
\end{document}